\documentclass[11pt]{article}
\usepackage{fullpage}

\RequirePackage{amsthm,amsmath}
\RequirePackage{natbib}
\RequirePackage[colorlinks,citecolor=blue,urlcolor=blue]{hyperref}
\usepackage{subfigure}
\usepackage{fullpage}
\usepackage{amsmath}
\usepackage{amssymb}
\usepackage{amsthm,graphicx,bm}

\usepackage{natbib}

\newcommand{\bTheta}{\boldsymbol{\Theta}}

\newcommand{\bD}{\boldsymbol{D}}
\newcommand{\bX}{\boldsymbol{X}}
\newcommand{\bY}{\boldsymbol{Y}}
\newcommand{\bW}{\boldsymbol{W}}

\begin{document}
\title{Modelling collinear and spatially correlated data}
%\author{}
\author{Silvia Liverani\\ Department of Mathematics, Brunel University London, UK\\ Medical Research Centre Biostatistics Unit, Cambridge, UK\\ Department of Epidemiology and Biostatistics, Imperial College London, UK  \and Aurore Lavigne\\ University of Lille 3, France \and Marta Blangiardo \\MRC-PHE Centre for Environment and Health, \\Department of Epidemiology and Biostatistics, Imperial College London, UK}
\date{}
\maketitle

\begin{abstract}
In this work we present a statistical approach to distinguish and interpret the complex relationship between several predictors and a response variable at the small area level, in the presence of i) high correlation between the predictors and ii) spatial correlation for the response. 

Covariates which are highly correlated create collinearity problems when used in a standard multiple regression model. Many methods have been proposed in the literature to address this issue. A very common approach is to create an index which aggregates all the highly correlated variables of interest. For example, it is well known that there is a relationship between social deprivation measured through the Multiple Deprivation Index (IMD) and air pollution; this index is then used as a confounder in assessing the effect of air pollution on health outcomes (e.g. respiratory hospital admissions or mortality). However it would be more informative to look specifically at each domain of the IMD and at its relationship with air pollution to better understand its role as a confounder in the epidemiological analyses.

In this paper we illustrate how the complex relationships between the domains of IMD and air pollution can be deconstructed and analysed using profile regression, a Bayesian non-parametric model for clustering responses and covariates simultaneously. Moreover, we include an intrinsic spatial conditional autoregressive (ICAR) term to account for the spatial correlation of the response variable.

\vspace{0.3cm}

\end{abstract}

\section{Introduction}
In many statistical applications a common challenge arises when trying to assess meaningful relationships between explanatory variables and outcomes through regression models, due to the potential collinearity of the explanatory variables. This issue is well known in epidemiological or social studies, for instance where questionnaires or surveys collect information on a large number of potential risk factors for particular end points; in this context a simplistic approach consists in examining  each variable in turn to avoid the instability in the estimates due to the collinearity, making it impossible to judge the more realistic complex relationship involving several risk factors at the same time. A different approach combines all the relevant variables into summary scores or indexes and assesses the relationship of these with the outcome of interest, which is free from the collinearity issue, but loses information on the single variables included in the summary. 

Recently, Dirichlet process mixture models have been used as an alternative to regression models \citep{Dunson08,Bigelow09}. In this paper we focus on the model known as profile regression and proposed by \cite{Molitor10}. Profile regression is a Bayesian non--parametric method which assesses the link between potentially collinear variables and a response through cluster membership. This allows to formally take into account the correlation between the variables without the need to create a summary score, giving more flexibility to the inferential process. Profile regression has been used on several applications in environmental and social epidemiology and the R package PReMiuM \citep{Liverani15} makes it readily available to any applied researcher. For instance \cite{Molitor10} considered the National Survey of Children's Health and in particular investigated a large number of health and social related variables on mental health of children age 6--17, while \cite{Papathomas11} focussed on profiles of exposure to environmental carcinogens and lung cancer in the EPIC European cohort.  Profile regression has also been used in environmental epidemiology \citep{Pirani15}, for studying risk functions associated with multi-dimensional exposure profiles \citep{Hastie13,Molitor14} as well as for looking for gene-gene interactions \citep{Papathomas12}.%Finally, \cite{Molitor11} used again the National Survey of Children's Health  to study the link between air pollution and poverty in Los Angeles and how this affects mental health of children living in that area.

In its present formulation, profile regression has only been used for studies based on cohorts or surveys where information on the predictors/outcomes is available on each individual; in this paper we extend the method to fit small area studies,  commonly used in epidemiological surveillance (see for instance \citealt{Elliott04}) or in studies where the interest lies on the spatial variability of an outcome \citep{Barcelo09} or on cluster detection \citep{Abellan08,Li12}. In this types of studies information is available at the area level rather than at the individual level and space is used as a proxy for any unmeasured variable; the common assumption is that areas which are close to each other are more similar than those further apart, suggesting that an additional source of correlation, namely \textit{spatial correlation} needs to be accommodated in the models. We incorporate it in the model  through a conditional autoregressive structure \citep{Besag91} based on a neighborhood definition, thus assuming that conditional on the neighborhood structure, two areas are independent from each other if they do not share boundaries. We apply the \textit{spatial profile regression} to the problem of environmental and social inequalities in London, jointly modelling social deprivation and air pollution to highlight the presence of environmental justice.

The paper is structured as follows. In Section \ref{sect:casestudy} we present the motivating example for our methodological development of the spatial profile regression, introducing the context of social and environmental inequalities and how they are related; we also describe the available data. In Section \ref{sect:methods} we provide a brief summary of the profile regression and present how to extend it to include spatial correlation.  In Section \ref{sect:results} we illustrate how the model works on evaluating the relationship between social deprivation and air pollution. Section \ref{sect:conclusion} presents some discussion points and ideas for future work. 
 
\section{Example: social deprivation and air pollution in London}\label{sect:casestudy}

The scientific literature reports mixed evidence on the link between socio-economic status and air pollution. Recent studies indicated that air pollution tends to influence most deprived groups, suggesting that people with lower socio-economic status are more likely to live in a more hazardous and polluted living environment, accidentally or deliberately \citep{Brown95,Oneill03,Blowers94,Morello02}. In particular, ecological studies using small areas such as neighbourhoods, census tracts and post codes, report this association, while studies carried out at a lower spatial resolution (e.g. region, country), thus characterised by more aggregate measurements of socio-economic characteristics, showed either non-existent or negative associations \citep{Davidson00,Laurent07}, presumably due to the large within-area variability not taken into account, or even an inverse association, with higher exposures in less deprived groups \citep{Perlin95}. In the UK several studies reported positive or non-linear correlation between environmental pollution and the deprivation index at both small area level and country level. However the results varied depending on the selection of environmental hazards and scale of analysis (Briggs et al., 2008), calling for some more research on the topic.

Understanding environmental and social inequalities is a key issue as growing health disparities appear between people with socially disadvantaged and privileged social classes, which can translate into increased mortality or morbidity for the low socio-economic groups across a wide range of diseases \citep{Brulle06,Benach01}, including lung cancer \citep{Pope11}, cardiovascular events \citep{Tonne07,Peters04}, and childhood respiratory diseases \citep{Morgenstern07}. 

In addition, the exposure of air pollution can lead to negative health outcomes acutely or chronically \citep{Chen08}. Previous studies reported possible mechanisms to explain how environmental exposures result in greater health impact among socially disadvantaged groups, who may have increased susceptibility to the effect of these exposures because of limited access to health care and psychosocial stress; underlying health conditions such as cardiovascular diseases and respiratory diseases that increase susceptibility to the effect of these exposure may also vary between deprived and privileged populations \citep{Oneill03,Morello06}. These environmental exposure inequalities are increasingly considered as a potential determinant of health disparities \citep{Morello06}. In addition, it has been suggested that the disparities grow in more deprived areas as health improves faster in high socio-economic groups \citep{Leyland07,Higgs98}. 

Although individual determinants (such as smoking) or individual risk responses (such as closing windows to avoid exposure)  may frequently contribute to these health inequities, only a fraction of the overall disparities are attributed to individual factors \citep{Lantz01}. In fact, human health is not only influenced by individual health behaviours but also by contextual and ecological factors \citep{Marmot07}. Furthermore, socio-economic status plays a potential role of confounding or effect modification in epidemiological studies investigating the relationship between environmental variables and health outcomes, especially at  aggregated level \citep{Blakely00,Blakely04}. The further effect of confounding and effect modification will potentially lead to bias of the results, whose level depends on the relationship between environmental pollutions and socio-economic status. Hence it is extremely important to study this association, which, at the moment still remains uncertain and subjected to the fundamental methodological issue of correlation between variables. 

To study this relationship in the present work we consider the following data:
\begin{itemize}
\item nitrogen oxides (NO$_x$), which is generated mainly through combustion, thus is a good proxy for traffic related air pollution. The data were obtained from the environmental research group at Kings College as annual mean for the period 2003-2010 at the Lower Super Output Area geographical level in Greater London (LSOA, 4,767 in Greater London) as part of the TRAFFIC project (\url{http://www.kcl.ac.uk/lsm/research/divisions/aes/research/ERG/research-projects/traffic/index.aspx}).

\item Index of Multiple Deprivation (IMD), publicly available from the Department for Communities and Local Government (\texttt{data.gov.uk}). It is commonly used at the small area level to synthesize multiple aspects of deprivation. It is originally built at LSOA level and is formed by 38 indicators collapsed into seven domains: Income, Employment, Health, Education, Crime, Access to Services (Housing) and Living Environment. As we want to evaluate the relationship between the domains of the IMD and air pollution (NOx) we have not considered the living environment domain, which includes air quality. IMD is available for 2004, 2007 and 2010 and we have considered the most recent one in this work (correlation between Index at different years ranges from 0.94 and 0.97).
\end{itemize}

\begin{figure}[ht]
\centering
\subfigure[NO$_x$ concentration for 2003-2010]{\includegraphics[scale=0.4]{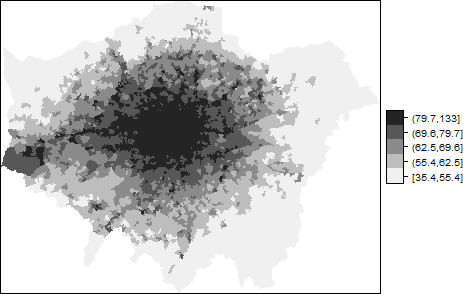}}
\qquad \subfigure[IMD score, 2010]{\includegraphics[scale=0.4]{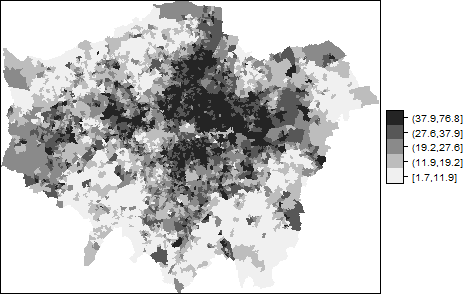}}
\caption{Quintilesof the NO$_x$ concentration (average 2003-2010) and of IMD score (2010) at LSOA level in Greater London.\label{Fig:NOX-IMD}}
\end{figure}

\begin{figure}[h]\centering
\subfigure[Income]{\includegraphics[scale=0.4]{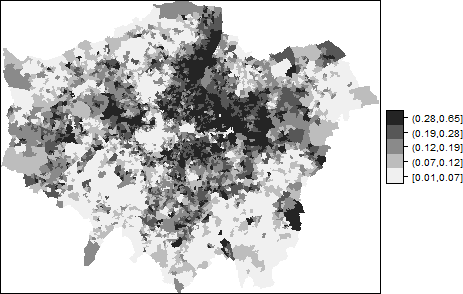}}\qquad
\subfigure[Employment]{\includegraphics[scale=0.4]{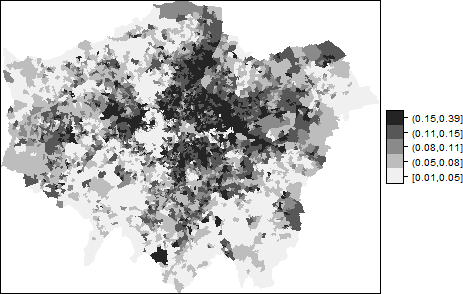}}

\subfigure[Crime]{\includegraphics[scale=0.4]{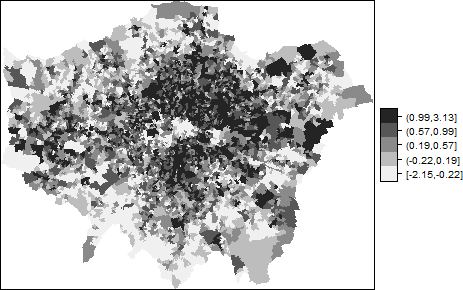}}\qquad
\subfigure[Barriers to Housing and Services]{\includegraphics[scale=0.4]{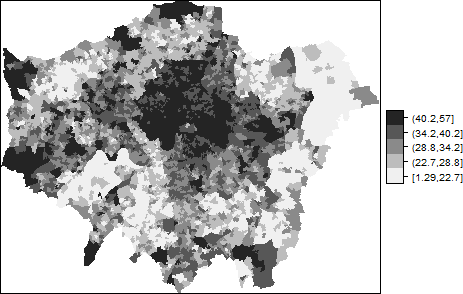}}

\subfigure[Education]{\includegraphics[scale=0.4]{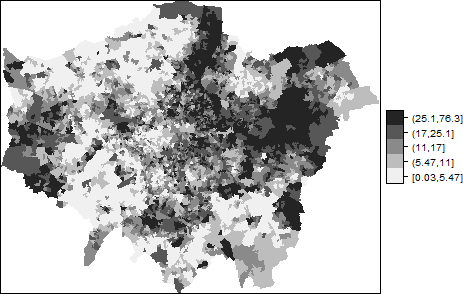}}\qquad
\subfigure[Health and Disability]{\includegraphics[scale=0.4]{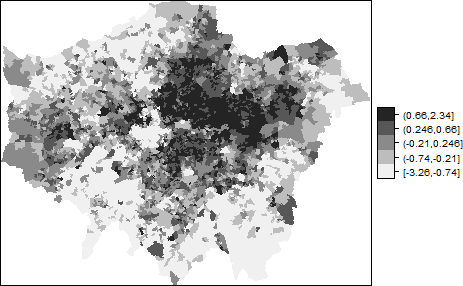}}
\caption{Maps of the six IMD domains considered: quintiles of the scores (note that higher positive values means higher deprivation.\label{Fig:IMD}}
\end{figure}

Figure \ref{Fig:NOX-IMD} shows the map for NO$_x$ (left) and IMD score (right) and a clear spatial pattern is visible in both: air pollution concentration increases steadily going from outer to inner London, while IMD shows the highest deprived areas in the northeastern part of London and most of the central southern part. However looking at the maps of each of the six domains highlights a different picture (Figure \ref{Fig:IMD}): Crime shows the absence of a clear pattern, with scattered areas of high crime (dark grey) next to areas of low crime (light grey); on the other hand income, employment and health/disability are in agreement with the total IMD score, while barriers to housing and services are more pronounced in central London and education shows more deprivation in East London. This suggests how simplistic is the approach that considers the total IMD score and highlights the importance of including all the domains to disclose the relationship between social characteristics and environmental pollution at a small area level. 

If we want to investigate the relationship between each domain and air pollution we cannot include all the domains in a regression model due to their  collinearity issues: the pairwise Pearson correlation between domains (Table \ref{tab:cor}) shows high values for income and employment (0.91), income and health (0.77), income and education (0.68), employment and health (0.81) and employment and education (0.64).

\begin{table}[ht]
\caption{Correlation between IMD domains. In bold correlation higher than 0.6. \label{tab:cor}}
\centering
\begin{tabular}{l|llllll}
\hline
	&      Income & Employ. &   Health & Educ. &  Hous.  &   Crime\\
\hline
Income  &   1.0 & \textbf{0.91} & \textbf{0.77} & \textbf{0.68} & 0.48 & 0.52\\
Employ. & 0.91 & 1.0 & \textbf{0.81} & \textbf{0.64} & 0.42 & 0.53\\
Health    & 0.77 & 0.81 & 1.0 & 0.55 & 0.41 & 0.59\\
Educ. & 0.68 & 0.64 & 0.55 & 1.0 & 0.16 & 0.36\\
Hous.   & 0.48 & 0.42 & 0.41 & 0.16 & 1.0 & 0.29\\ 
Crime    &  0.52 & 0.53 & 0.59 & 0.37 & 0.29 & 1.0\\
\hline
\end{tabular}
\end{table}

\clearpage
\section{Modelling highly correlated covariates with profile regression}\label{sect:methods}
To include all the domains into the same statistical model we use the profile regression, a model that non-parametrically links a response vector $\bY$ to covariate data $\bX$ through cluster membership. It was proposed by \cite{Molitor10} and it has been implemented in the R package PReMiuM \citep{Liverani15}. 

Profile regression implements a Bayesian clustering model through a Dirichlet process mixture model. The data $\bD=(\bY,\bX,\bW)$ contain the response $\bY$, covariate $\bX$ and fixed effects $\bW$ if they are available. The fixed effects are potentially confounding variables. In our running example the response is the nitrogen oxides, the covariates are the selected six domains of IMD and we do not include fixed effects. 

For each individual $i$, for $i=1,\ldots,n$, the response is given by $y_i$, the covariate vector $\mathbf{x}_i$ and the fixed effect vector $\mathbf{w}_i$. The data are then jointly modelled as the product of a response model and a covariate model, leading to the following likelihood:
\begin{equation*}
f(\mathbf{x}_i,y_i | \bTheta_{Z_i},\boldsymbol\Lambda,W_i,\psi) = \sum_c \psi_c f(\mathbf{x}_i | z_i = c,\phi_c)f(y_i|z_i=c,\theta_c,\boldsymbol\Lambda,\mathbf{w}_i) 
\end{equation*}
where $z_i = c$, the allocation variable, indicates that individual $i$ belongs to cluster $c$. The parameters $\boldsymbol\Lambda=(\boldsymbol{\theta},\boldsymbol\phi)$ are cluster specific and represent the contribution of the response and the covariates to the mixture model. There is also the possibility to include additional fixed effects $\mathbf{w}_i$ for each individual, which are constrained to only have a global (i.e., non-cluster specific) effect on the response $y_i$. The parameters $\psi$ are the mixture weights. 

Multicollinearity arises when regression models of the response with respect to highly correlated covariates are implemented, due to identifiability issues. However, as here the response is conditionally independent from the covariates, we do not encounter such issues, but we can explore in depth the potentially complex relationship between response and covariates. 

The prior model for the mixture weights is given by the stick-breaking priors (constructive definition of the Dirichlet process), that is,
\begin{eqnarray*}
\psi_c &=& V_c \prod_{l<c} (1-V_l) \quad \mbox{for all $c$},\\
\psi_1 &=& V_1,\\
 V_c &\sim& \mbox{Beta} (1,\alpha) \quad \mbox{i.i.d.}
\end{eqnarray*}
The parameter $\alpha$ can be fixed or can have a Gamma$(s_\alpha,r_\alpha)$ distribution with $s_\alpha$ and $r_\alpha$ as shape and rate parameters respectively. Other prior models for the mixture weights are possible, and, for example, the Pitman-Yor construction is also available in the R package PReMiuM. 

The covariate model $f(\mathbf{x}_i | z_i = c,\phi_c)$ can be defined as continuous or discrete. In the continuous case $\bX$ assumes a mixture of Gaussian distributions. In the discrete case for each individual $i$, $\mathbf{x}_i$ is a vector of $J$ locally independent discrete categorical random variables, where the number of categories for covariate $j=1,2,\ldots,J$ is $K_j$. Then we can write $\Phi_c=(\Phi_{c,1},\Phi_{c,2}\ldots,\Phi_{c,J})$ with $\Phi_{c,j} = (\phi_{c,j,1},\phi_{c,j,2},\ldots,\phi_{c,j,K_j})$ and
\begin{equation}
   f(\mathbf{x}_i | z_i = c,\phi_c)=\prod_{j=1}^J\phi_{{Z_i},j,X_{i,j}}. 
\end{equation}
We let $a = (a_1 , a_2 ,\ldots , a_J )$, where for $j = 1, 2,\ldots , J$, $a_j = (a_{j,1} , a_{ j,2} ,\ldots , a_{ j,K_{ j}} )$ and $\Phi_{ c,j} \sim \mbox{Dirichlet}(a_j )$. The covariate model can also be defined as a mixture of continuous and discrete covariates.  

The response model $f(y_i|z_i=c,\theta_c,\boldsymbol\Lambda,\mathbf{w}_i) $ can be defined as binary, categorical, count (modelled as Binomial or Poisson) or Gaussian. For example, for Gaussian response the mixture model is extended to contain $\theta_c$ for each $c$ and the global parameters $\boldsymbol\Lambda=(\boldsymbol\beta, \sigma^2_Y)$. These parameters allow us to write the response model as:
\[
f(y_i|z_i=c,\theta_c,\boldsymbol\Lambda,W_i)=f(y_i|z_i=c,\theta_c,\boldsymbol\beta,\sigma^2_Y,W_i)=\frac{1}{\sqrt{2\pi\sigma^2_Y}}\exp\left\{-\frac{1}{2\sigma^2_Y}(Y_i-\lambda_i)^2\right\},
\]
where $\lambda_i=\theta_{Z_i}+\boldsymbol\beta^\top W_i$ and $\boldsymbol\beta$ represent the effect of the counfounding variables, the fixed effects, on the response. For each cluster $c$, we adopt a $t$ location-scale distribution on $\theta_c$, with hyperparameters $\mu_\theta$ and $\sigma_\theta$ with 7 degrees of freedom. Similarly, we adopt the same prior for the fixed effect $\beta_k$, but with hyperparameters $\mu_\beta$ and $\sigma_\beta$. We set $\tau_Y = 1/\sigma_{Y}^2$ to Gamma$(s_{\tau_Y},r_{\tau_Y})$, where $s_{\tau_Y}$ and $r_{\tau_Y}$ are the shape and rate hyperparameters. 

More details on the Markov chain Monte Carlo (MCMC) algorithm for this model are provided in \cite{Liverani15}. When the signal in the data is strong the MCMC results for different runs, with different initial values and chain lengths, give stable results. However, this is not the case when the signal is not strong. \cite{Hastie15} discuss strategies to identify convergence issues. They recommend starting the MCMC with a large number of clusters as the algorithm can struggle to explore the partition space when starting with a small number of clusters if the signal is not strong. When many clusters are identified it is more challenging to characterise each cluster meaningfully and interpretation of the results can be facilitated by the posterior predictive distrbution. Moreover, they suggest the use of the posterior distribution of predictive profiles for the assessment of convergence instead of the posterior distribution of the parameters. This is because the posterior distribution of the parameters can appear to have converged when the model as a whole has not (often the case for $\boldsymbol\beta$), or cannot be used for this scope because they are cluster-specific and the number of clusters changes at each iteration (such as $\theta_c$).
 
It is often useful to characterise the partition which is most supported by the data. However, as at each iteration of the sampler individual profiles are assigned to clusters, the MCMC output is very rich. \cite{Molitor10} developed methods to process this output to make useful and interpretable inference.  Several methods for this are available in the R package PReMiuM but we find the most robust method is to process the similarity matrix using partitioning around medoids (PAM), which is available in the R package cluster. First of all, a score matrix is constructed, where each element of the matrix is set equal to 1 if individuals $i$ and $j$ belong to the same cluster and 0 otherwise. Then a similarity matrix $S$ is computed by dividing each element of the score matrix by the number of iterations, so that $S_{ij}$  denotes the probability that individuals i and j are assigned to the same cluster. PAM then assigns individuals to clusters in a way consistent with matrix $S$. 

%The R package PReMiuM also includes two methods for variable selection \citep{Papathomas12} and postprocessing functions to identify the optimal partition in the rich MCMC output.  

\subsection{The Spatial Conditional Autoregressive Model}

When clustering data from small area studies, we need to modify the model to account for spatial correlation. In this paper we propose to extend the response model described above to include an intrinsic spatial conditional autoregressive (ICAR) term \citep{Besag91} as follows. The likelihood component for the Gaussian response becomes 
\[
f(y_i|z_i=c,\theta_c,\boldsymbol\Lambda,W_i)=f(y_i|z_i=c,\theta_c,\boldsymbol\beta,\sigma^2_Y,u_i,W_i)=\frac{1}{\sqrt{2\pi\sigma_Y^2}}\exp(-\frac{1}{2\sigma_Y^2}(Y_i-\lambda_i)^2)
\]
where $\lambda_i= \theta_{Z_i}+  W_i\boldsymbol\beta + u_i $ and $u=(u_1, \cdots, u_n) \sim N(0, \mathbf{\tau P})$ with $\mathbf{P}=\{P_{ij}\}$ a precision matrix such that 
\[
P_{ij} = \begin{cases} 
n_i & \mbox{if} \;  i = j\\
-I\{i\sim j\}  &\mbox{if} \; i \ne j \\
\end{cases} 
\]
where $n_i$ is the number of neighbours of subject $i$, $I$ is the indicator function and $i \sim j$ indicates that regions $i$ and $j$ are neighbours. The prior of $\tau$ is given by 
\[
\tau \sim \mbox{Gamma}(a_{\tau}, b_{\tau})
\]
such that 
\[
E(\tau)=\frac{a_{\tau}}{b_{\tau}} \quad and \quad \mbox{Var}(\tau)=\frac{a_{\tau}}{b_{\tau}^2}.
\] 
Details of the sampling strategy for the ICAR parameters are given in Appendix \ref{app1}. We have implemented this model in the R package PReMiuM for Gaussian and Poisson responses. 

\section{Results}\label{sect:results}

We have fit profile regression to the data using the R package PReMiuM \citep{Liverani15}. The pollution data are modelled with a Gaussian distribution including a spatial ICAR term. The covariate profiles, given by the selected IMD domains, are modelled with a discrete distribution, as we have transformed each IMD domain into quintiles. We have not included any additional fixed effects. The results were very robust on several MCMC runs using a range of initial values and different chain lengths. We present here the results obtained with 5,000 iterations after a burn in of 5,000 with the following hyperparameter settings. 
\begin{eqnarray*}
&&s_\alpha = 2, \quad \quad r_\alpha=1,\\
&&a_1=\ldots=a_6=1,\\
&&\mu_\theta=0, \quad \sigma_\theta=2.5,\\
&&\mu_\beta=0, \quad \sigma_\beta=2.5,\\
&&s_{\tau-Y}=2.5, \quad r_{\tau-Y}=2.5.
\end{eqnarray*}

\begin{figure}[h]
\centering
\includegraphics[height=7cm]{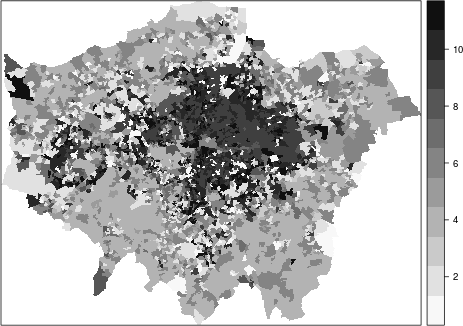}
\caption{Geographical representation of the eleven clusters of the areas in Greater London identified by profile regression. The colours reflect the mean of the observed pollution levels, with dark grey identifying the most polluted clusters and light grey the least polluted clusters. \label{fig:clusters}}
\end{figure}

In this section we aim to illustrate how profile regression can help shed light on complex patterns between highly covariates covariates and response. We propose several ways to explore the results.  

The main output of profile regression are the clusters, given by regions with similar covariate and response characteristics. We have used different types of plots, which highlight specific features, at different level of details. In Figure \ref{fig:clusters} we represent the clusters geographically. The eleven clusters identified are plotted using colours that reflect their observed pollution levels. As expected, the most central areas have higher pollution levels. In particular we see that areas that belong to clusters with higher observed pollution levels are mostly located in North-East London, where areas with higher levels of deprivation are found. In contrast, the less deprived areas in South-West London are clustered together and have lower mean for the observed pollution levels.

\begin{figure}[!htbp]
\centering
\includegraphics[angle=90,height=18cm]{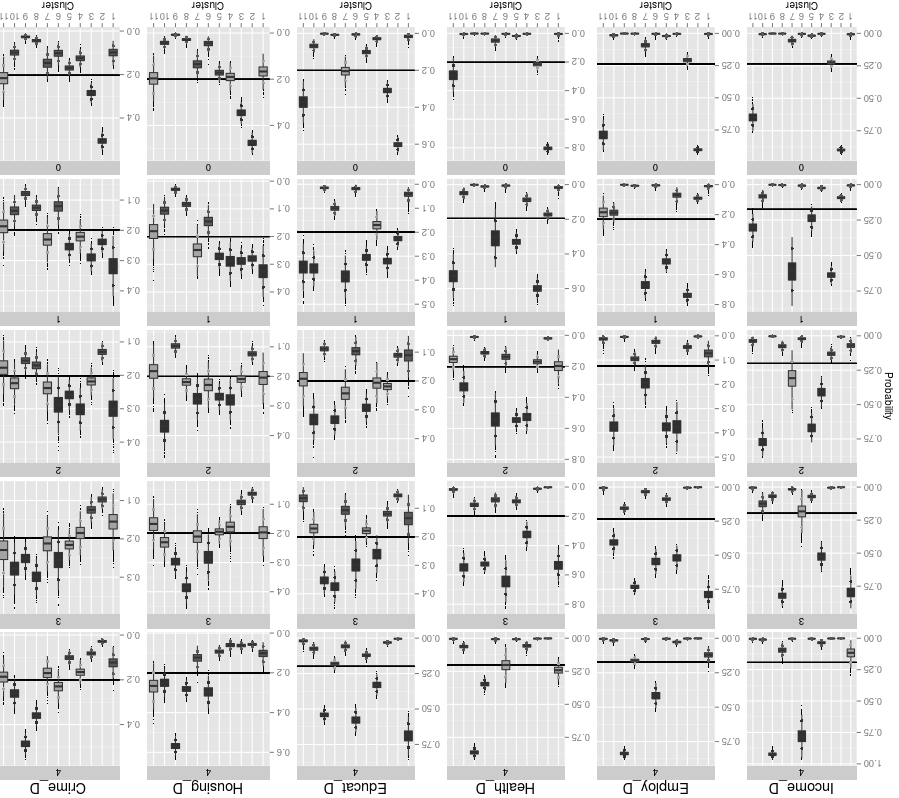}
\caption{Summary plot of the posterior distribution of the parameters $\Phi_c$, for $c=1,\ldots11$. Each column $j$ in the figure represents $\Phi_{c,j,k}$, for $c=1,\ldots,11$ and $k=0,1,\ldots,4$. Within a column $j$, each row $k$ is a visualisation of the boxplots for $\phi_{c,j,k}$ for each cluster $c$.\label{fig:summary}}
\end{figure}

The cluster data are high-dimensional and complex. Figure \ref{fig:summary} provides a visual representation of the parameters $\Phi_c$. Through the boxplots of the MCMC samples of the parameters $\Phi_c$, this figure provides also a representation of the uncertainty around these parameters. Each column $j$ in the figure represents $\Phi_{c,j,k}$, for $c=1,\ldots,11$ and $k=0,1,\ldots,4$ (the five quintiles of each covariate). Within a column $j$, each row $k$ is a visualisation of the boxplots for $\phi_{c,j,k}$ for each cluster $c$. This visual representation provides a further insight into the eleven clusters. We can identify patterns in the relationships between pollution and IMD domains at a glance. For example, Housing and Crime appear to generally increase as pollution levels increase, while the other domains show less linear patterns. We can also see here the details of the distributions of the levels of covariates that define the different clusters. Cluster 2 has the lowest levels of deprivation, although not the lowest levels of mean pollution. In contrast, cluster 9 has the highest levels of deprivation, and high levels of pollution, although not the highest among all clusters. The mean IMD per cluster highlights the complex relationship between IMD, pollution and deprivation. For example, for cluster 6, which has a high mean IMD, there is strong deprivation for the first four domains. However, the domains of Crime and Housing are rather evenly spread among all levels of the covariates, suggesting that they do not contribute to the deprivation that characterises these areas. This is an example of a complex pattern that cannot be identified when the domains are simply collapsed into the IMD. 

\begin{figure}[!htbp]
\centering
\includegraphics[height=12cm]{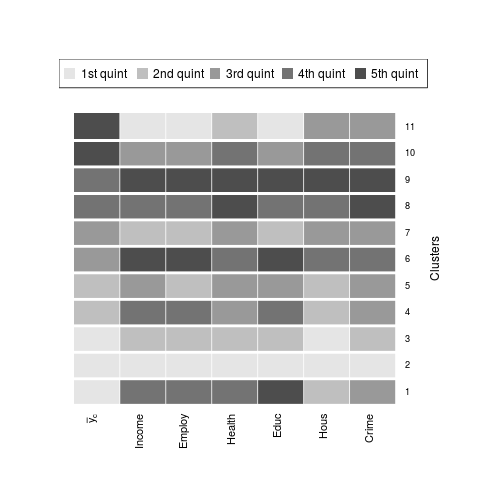}
\caption{Summary table of the clusters. The quintiles for pollution and each domain of the IMD are shown for each cluster. \label{fig:heatmap}}
\end{figure}

In Figure \ref{fig:heatmap} we provide a summary of the posterior means for each cluster. Each row represents a cluster. The columns represents, respectively, the mean observed pollution and each domain of IMD. The colour of each column of the matrix corresponds to a quintile of the distribution of that variable. As before, the clusters are ordered by their observed pollution level. Note that the colours in the matrix do not become darker (or lighter) in a smooth manner. Together, Figure \ref{fig:heatmap} and Figure \ref{fig:clusters} suggest that the areas of low pollution and low deprivation are in outer London. As we get closer to the centre, pollution increases and many of the deprivation variables increase levels. However, there are many notable exceptions to this. For example, cluster 11 has the highest levels of pollution, but among the lowest levels of deprivation. On the contrary, cluster 1 has the lowest level of pollution, but rather high levels of deprivation on all domains except Housing. 

\begin{figure}[!htbp]
\centering
\includegraphics[height=6cm]{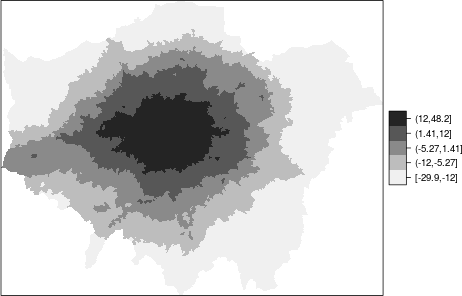}
\caption{Posterior mean of the spatial conditional autoregressive term.\label{fig:u}}
\end{figure}

Figure \ref{fig:u} shows the posterior mean of the spatial term $\exp(u_i)$ for each area, which accounts for the residual spatial variation in NOx after having adjusted for the cluster assignment. The map presents a clear pattern going from central London (darker) to outer London (lighter) with values ranging from -30 to 48 $\mu g/m^3$, thus suggesting that the model picks up the spatial dependence in air pollution concentration which is not explained by deprivation.

\begin{figure}[!htbp]
\centering
\includegraphics[height=8cm]{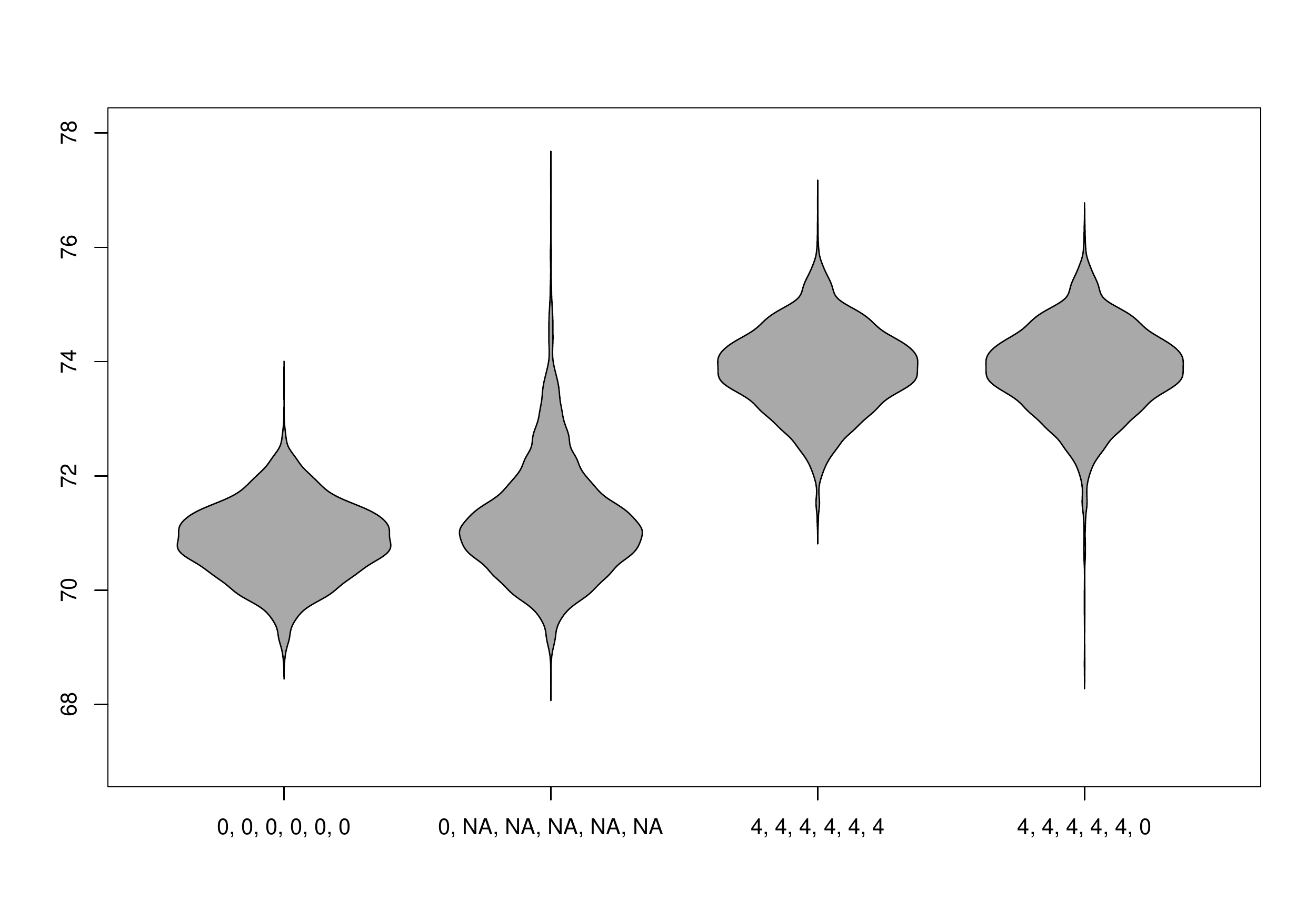}
\caption{Beanplots of the posterior predictive distributions for these four pseudo profiles: $(0,0,0,0,0,0), (0,\text{NA},\text{NA},\text{NA},\text{NA},\text{NA}), (4,4,4,4,4,4), (4,4,4,4,4,0)$ where the elements of each vector represent respectively the IMD domains (Income, Employment, Health, Education, Housing, Crime). \label{fig:preds1}}
\end{figure}

We can explore the relationships between covariates and response further by looking at the posterior predictive distributions. The profile regression model allows us to predict the pollution level for specific combinations of the IMD domains. If we wish to understand the role of a particular covariate or group of covariates, we can specify a number of predictive scenarios (pseudo-profiles), that capture the range of possibilities for the covariates that we are interested in \citep{Hastie13}. For each of these pseudo-profiles we can see how these would have been allocated in our mixture model to understand the level of pollution associated with them once we have accounted for the spatial residuals. 

In Figure \ref{fig:preds1} we show beanplots of four pseudo-profiles: $(0,0,0,0,0,0)$, $(0,\text{NA},\text{NA},\text{NA},\text{NA},\text{NA})$, $(4,4,4,4,4,4)$, $(4,4,4,4,4,0)$. The elements of each vector represent the IMD domains in the following order: Income, Employment, Health, Education, Housing, Crime. Each of these beanplots shows the pseudo-profile corresponding to particular values of the IMD domain of interest for an average area (ie. including the mean spatial residual, which is 0). When `NA' is set, this allows that domain to vary, de facto marginalising for it, i.e. capturing all possible values it can take. For example, the beanplot for the first pseudo-profile on the left presents the posterior predictive distribution of NOx for areas with the lowest levels of deprivation and shows values between 68 and 75$\mu$g/m$^3$. Using this as benchmark we can make some comparisons: areas characterised by low Income and marginalising for the remaining domains (second beanplot from the left) have a much wider posterior predictive distribution, with values going from 67 to 78$\mu$g/m$^3$ while areas with in the highest quintile of deprivation for all the domains (third beanplot from the left) present consistently highest level of pollution (ranging between 71 to 77$\mu$g/m$^3$). The last beanplot shows the posterior predictive distribution of NOx in an area where crime has decreased to the first quintile (for instance through the implementation of a policy) while the other domains remains in the last quintile. Comparing it with the previous one it an be seen a similar distribution, but with a lower tail which could be a consequence of the policy implementation. 

All predictive profiles can be computed and we provide here only these examples to show how these can be interpreted, if there was an interest in the posterior predictive distribution of specific combinations of deprivation levels, these could be explored in depth. Moreover, if there was interest in a specific area, the pseudo-profiles could be adjusted by adding the spatial residual. 

\section{Discussion}\label{sect:conclusion}

In this paper we have considered a spatially-correlated response variable and a set of highly correlated covariates. We have extended the profile regression model, a Bayesian clustering method used to deal with collinearity in the predictors, to account for spatial correlation adding a spatial conditionally autoregressive term. 

We have applied our method to explain the relationship between air pollution and social deprivation in Greater London. The Index of Multiple Deprivation is commonly used as a proxy for deprivation, as its domains usually cannot be analysed individually due to the high correlation between them. We have illustrated how profile regression can produce meaningful and useful results which shed light on the complex non linear relationship between pollution and the different domains.  

We want to stress that we are not framed in a standard regression approach, where the interest is to estimate the effect of each predictor on the outcome, as we do not attempt to explain the level of air pollution through the IMD domains. On the other hand through cluster assignment the profile regression  is able to disentangle the complex relationship between IMD and air pollution; this method has the added benefit of providing readily available prediction estimates which can be used to evaluate how the response could change for specific combinations of the predictors, and which could be used to evaluate the effect of policies.

A limitation of our model is that in its present formulation the spatial structure is not included on the cluster allocation, thus it accounts for local spatial dependency in the response, but not in the covariates, which is an extension we are going to work on in the future. 

%The code and the data to replicate our figures and results are available on the journal's webpage. 

\section*{Acknowledgements}
Marta Blangiardo acknowledges support from the UK NERC-MRC funded project “Traffic pollution and health in London” (NE/I00789X/1). Silvia Liverani acknowledges support from the Leverhulme Trust (ECF-2011-576). 

\appendix
\section{Sampling for the spatial ICAR parameters} \label{app1}
We include details of the sampling algorithm for the spatial ICAR parameters for Gaussian and Poisson distributed response. We have implemented both in the R package PReMiuM.

\subsection{Gaussian response}

The conditional distribution for $u_i$ is given by 
\begin{eqnarray*}
\log(p(u_i|u_{-i},\boldsymbol\beta,\boldsymbol\theta,\sigma_Y^2,Z_i,\tau,Y_i,W_i,T_i)) & \propto  & \log p(Y_i|u_i,\boldsymbol\beta,\theta,Z_i,\sigma^2_Y,W_i,T_i) + \log p(u_i|u_{-i},\tau)\\
& \propto & -\frac{1}{2\sigma_Y^2}(Y_i-(\theta_{Z_i}+  W_i\boldsymbol\beta + u_i))^2 -\frac{1}{2}\tau n_i (u_i-\bar{u}_i)^2\\
& \propto & -\frac{1}{2\sigma^2_i}(u_i - m_i)^2\\
\end{eqnarray*}
with
\[
\left\{\begin{array}{l}
m_i=\frac{\frac{1}{\sigma_Y^2}(Y_i-\theta_{Z_i}-W_i\boldsymbol\beta)-\tau n_i \bar{u}_i}{\frac{1}{\sigma_Y^2}+\tau n_i}\\
\sigma^2_i=\frac{1}{\frac{1}{\sigma_Y^2}+\tau n_i}
\end{array}\right.
\]
with , $\bar{u}_i=\frac{1}{n_i}\sum_{j \in \rho_i}u_j$ $\rho_i$ is the set of neighbours of $i$. Thus, for Normal response the prior is conjugated, the conjugated complete conditional distribution is Normal with mean $m_i$ and variance $\sigma_i^2$. The conditional distribution for $\tau$ is given by 
\[
\log(p(\tau|u))= (a_{\tau}+\frac{n-1}{2}-1)\log(\tau)-\tau(b_{\tau}+\frac{1}{2}u^T\mathbf{P}u)
\]
Thus $\tau \sim \mbox{Gamma}(a_{\tau}+\frac{n-1}{2},b_{\tau}+\frac{1}{2}u^T\mathbf{P}u)$. 

\subsection{Poisson response} For Poisson response, suitable for count data, the likelihood is given by 
\[
f_Y(y_i|z_i=c,\theta_c,\boldsymbol\Lambda,W_i)=p(Y_i|z_i=c,\theta_c,\boldsymbol\beta,u_i,W_i)=\frac{\mu_i^{Y_i}}{Y_i!}\exp\{-\mu_i\},
\]
where each individual $i$ is associated with an expected offset $E_i$, 
\[ 
\mu_i=E_i \exp\{\lambda_i\},\;\;\;\textrm{for } \lambda_i=\theta_{Z_i} + \boldsymbol\beta^\top W_i.
\]
As for the Gaussian response, the parameters $u=(u_1, \cdots, u_n) \sim N(0, \mathbf{\tau P})$ with $\mathbf{P}=\{P_{ij}\}$ a precision matrix such that 
\[
P_{ij} = \begin{cases} 
n_i & \mbox{if} \;  i = j\\
-I\{i\sim j\}  &\mbox{if} \; i \ne j \\
\end{cases} 
\]
where $n_i$ is the number of neighbours of subject $i$, $I$ is the indicator function and $i \sim j$ indicates that regions $i$ and $j$ are neighbours. The prior of $\tau$ is given by 
\[
\tau \sim \mbox{Gamma}(a_{\tau}, b_{\tau})
\]
such that 
\[
E(\tau)=\frac{a_{\tau}}{b_{\tau}} \quad and \quad \mbox{Var}(\tau)=\frac{a_{\tau}}{b_{\tau}^2}.
\] 

The conditional distribution for $u_i$ is given by 
\[
\log(p(u_i|u_{-i},\boldsymbol\beta,\theta,Z,\tau,Y))=Y_iu_i-E_i\exp(X_i\boldsymbol\beta+\theta_{Z_i}+u_i)-\frac{1}{2}\tau n_i (u_i-\bar{u}_i)^2
\]
with $\bar{u}_i=\frac{1}{n_i}\sum_{j \in \rho_i}u_j$ and $\rho_i$ is the set of neighbours of $i$. We implemented an adaptive rejection sampler for $u_i$. The conditional distribution for $\tau$ is given by 
\[
\log(p(\tau|u))= (a_{\tau}+\frac{n-1}{2}-1)\log(\tau)-\tau(b_{\tau}+\frac{1}{2}u^T\mathbf{P}u).
\]
Thus $\tau \sim \mbox{Gamma}(a_{\tau}+\frac{n-1}{2},b_{\tau}+\frac{1}{2}u^T\mathbf{P}u)$.

\bibliographystyle{chicago}
\bibliography{biblio}

\end{document}